# H-T phase diagram of the metamagnetic transition in $URu_2Si_2$, measured in high d. c. magnetic fields

A. Suslov[1,*], J. B. Ketterson[2], D. G. Hinks[3] and Bimal K. Sarma[1],

[1] Department of Physics, University of Wisconsin-Milwaukee, PO Box 413, Milwaukee, WI 53201

[2] Department of Physics and Astronomy, Northwestern University 2145 Sheridan Road, Evanston, IL 60208

[3] Materials Science and Technology Division, Argonne National Laboratory, Argonne, IL 60439

We have studied the ultrasonic velocity and ac susceptibility of $URu_2Si_2$ in d. c. magnetic fields up to 45T. A significant difference between the constructed H-T phase diagram and that extracted from earlier pulsed fields measurements can be explained in terms of a large magnetocaloric effect. A hysteresis of metamagnetic transition at low temperatures and a new phase separation line have been observed for the first time.

$URu_2Si_2$ is among the few heavy fermion compounds that show very interesting magnetic properties. Neutrons and X-ray scattering experiments [1-5] show that there is an antiferromagnetic transition at 17K in zero field [6-11] with the sublattice magnetization being along the c-axis. In addition, at high magnetic fields (35-40T) this compound shows a three-step metamagnetic transition [12-18]. Theoretical explanations of this three-fold splitting involve the crystalline anisotropy [18, 19].

Despite the fact that the antiferromagnetic transition has been investigated for 15 years, the nature of the phenomena is still unknown. There is no theory, which can quantitatively explain how the small low temperature magnetic moment $(0.03\pm0.01)\mu_B/U$ [1-5] which arises in $URu_2Si_2$ can have such a large effect on other physical properties, for example, specific heat [7-10], thermal expansion [10, 11], and resistivity [6, 9]. Some theoretical models [20-22 and references therein] qualitatively explain the behavior by invoking a *hidden order parameter*, whose origin still remains obscure.

In addition to difficulties with the theoretical understanding, various measurements of some important experimental properties are contradictory. Studies of the metamagnetic and antiferromagnetic transitions in $URu_2Si_2$ fall naturally into two groups. In the first group [6-8, 11] d. c. magnetic fields were used. To the best of our knowledge, until quite recently these measurements have been restricted to a maximum field of about 25T and therefore to investigations of antiferromagnetic transition only. In the second group pulsed magnetic fields were employed extending up to 60T [12-18]. Figure 1 shows the results of measurements from both groups; note the strong difference between the H-T phase diagrams extracted from the d. c. and pulsed field measurements, which we now discuss. All of the d. c. field studies show a single phase-separation line. This branch is an extension of the well-known zero-field transition, which occurs in $URu_2Si_2$ at about 17K. With increasing field, the transition temperature falls, reaching a value of about 10K at 25T. On the other hand, the H-T diagrams based on the pulsed fields investigations do not have any transition lines in common with the line emanating from ≈17K observed in the d.c. measurements. Instead, three temperature-independent branches are seen at about 35T, 37T and 39T in the temperature range 1 - 3K [12]. At 3K the middle transition disappears and another one appears at 36T. The three transitions (at 35T, 36T and 39T) are almost temperature independent and may be seen in the range 3 - 14K. At 14K (called point A in Fig. 1) the middle transition completely disappears and only two transitions (at 35T and 39T) survive up to 16K; however they remain temperature independent. Above 16K (point B) only one transition was observed in ref. [12] (it is not clear which of the two survives) while no transition was observed in ref. [13]. While only the temperature range up to 20K is shown in Fig. 1, the remaining transition continues up to about 40T for temperatures up to 60K.

---

[*] Permanent address: A. F. Ioffe Physical-Technical Institute, Russian Academy of Sciences, St.-Petersburg, Russia.

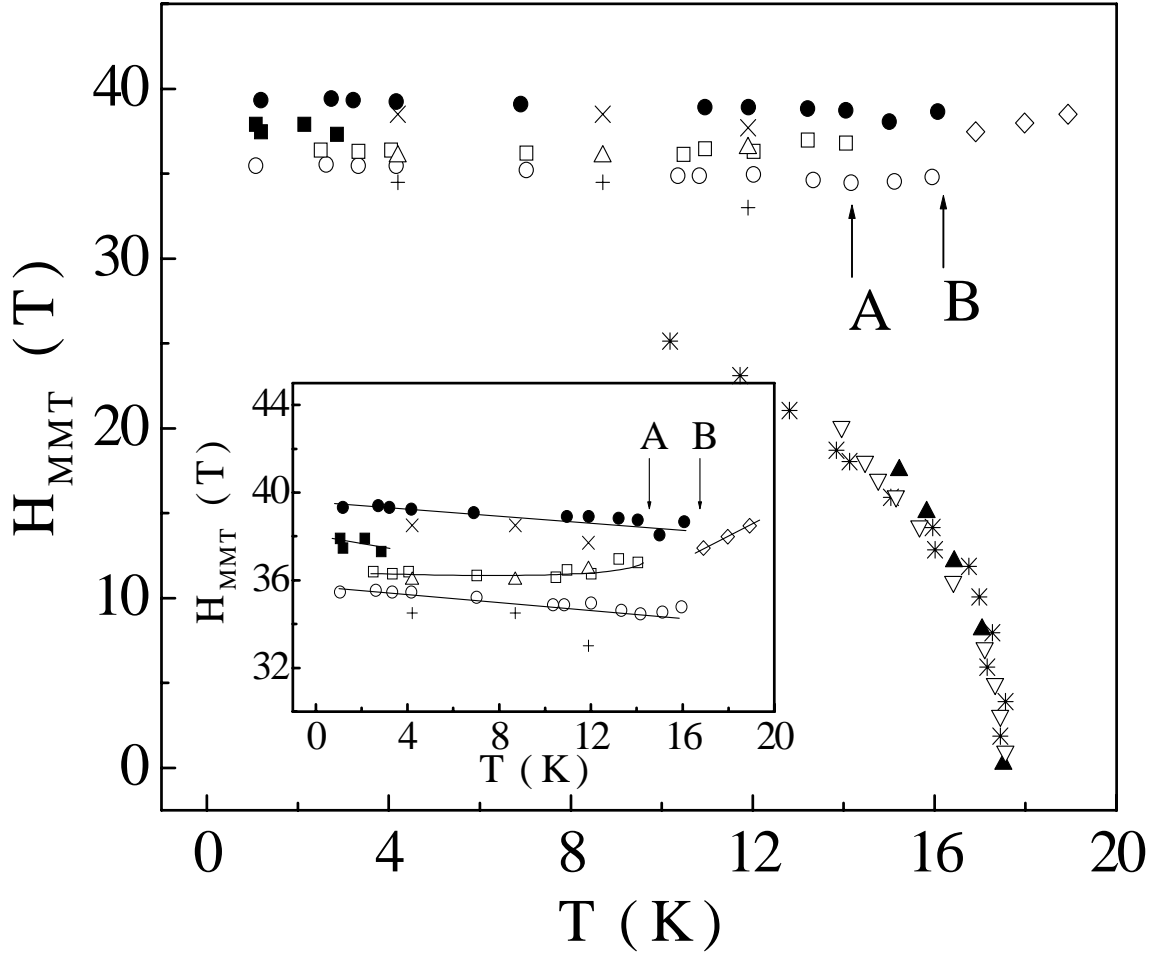

**Figure 1.** B-T phase diagram of the metamagnetic transition in $URu_2Si_2$ constructed from magnetoresistance: ✶ [6], specific heat: ▲ [7], and magnetization: ▽ [8] measurements in DC magnetic fields and combined together with magnetization: ○, ■, □, ●, ◇ [12] and ultrasound: +, △, × [13, $c_{11}$-mode] measurements in pulsed magnetic fields. Inset: Pulsed fields phase diagram, y-axis scale is increased in three times, the lines are guides to eye.

The difference between pulsed field and d. c. phase diagrams was briefly mentioned in [6, 12]. In both articles the difference was attributed to a different origin for the metamagnetic and the antiferromagnetic transitions. This explanation, and the fact that the branch observed in d. c. fields is not seen in pulsed fields, are quite surprising. In an effort to resolve this difference between the phase diagrams we used the hybrid magnet at the Tallahassee, FL site of National High Magnetic Field Laboratory to study sound propagation (Fig. 2) as well as the a. c. susceptibility and magnetocaloric effect (Fig. 3) in magnetic fields up to 45 T and in the temperature range 1–20K. The velocity of longitudinal sound propagated along the c-axis ($c_{33}$-mode) was measured at frequency 105MHz; the field was applied along the c-axis.

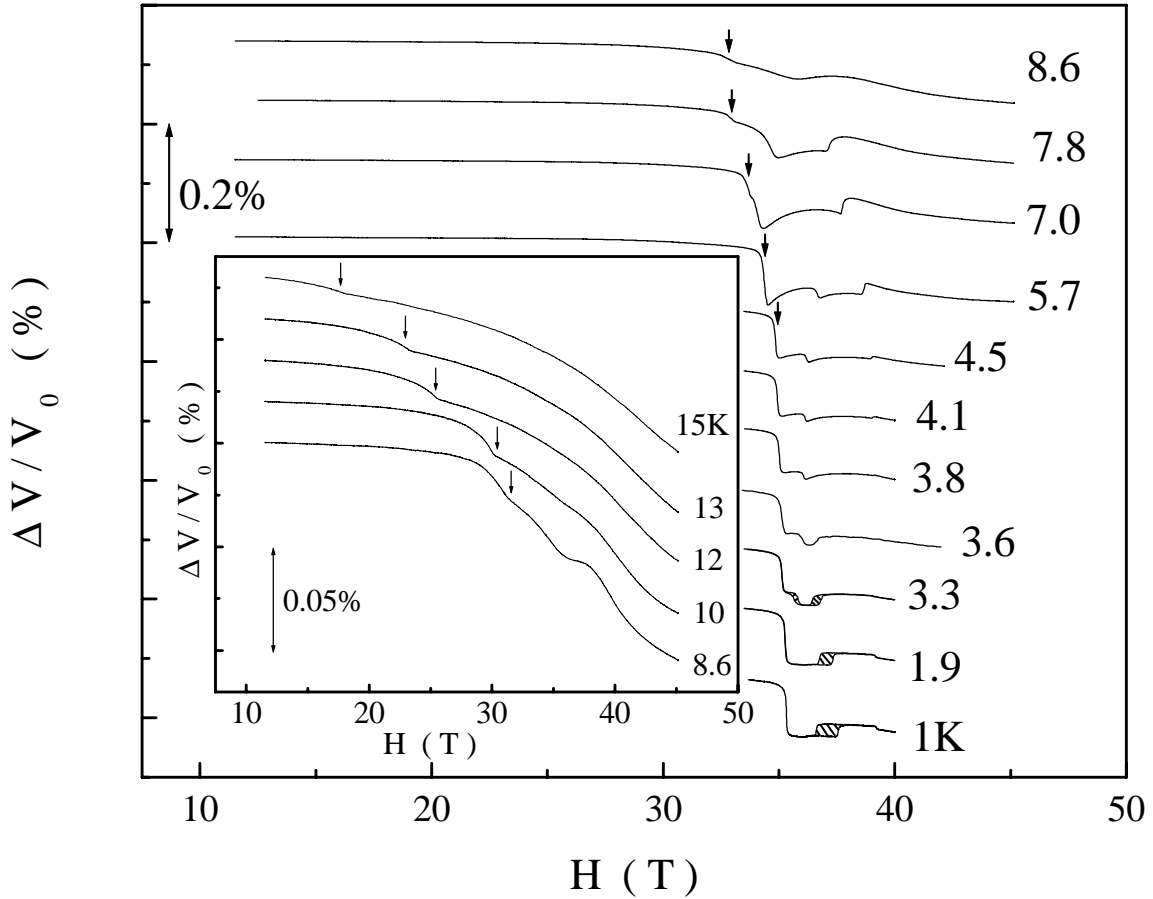

**Figure 2.** Magnetic field dependence of ultrasound velocity on $URu_2Si_2$ at different temperatures. The curves are shifted by about 0.1% for clarity. The arrows point transition 1. At lows temperatures a few curves show a hysteresis of the middle transition (36-37T). The hatched area shows the hysteresis region.

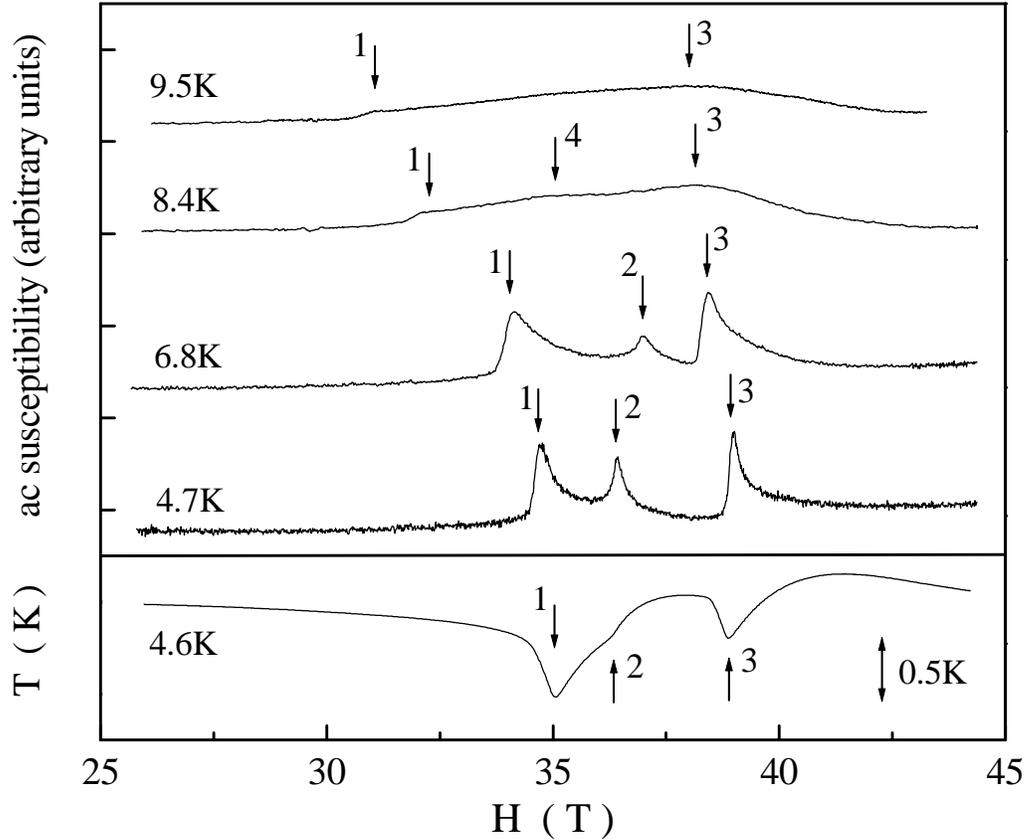

**Figure 3.** Magnetic field dependence of AC susceptibility voltage at different temperatures. The lowest curve is MCE trace (on field sweep up) starting at 4.6K.

We first discuss the ultrasonic results. As one can see from Fig. 2, three step-like changes occur in the magnetic field dependence of ultrasound velocity at a temperature of about 4.5K. The position of the velocity steps coincides with the H-value of the transitions measured in [12] at this temperature. The H-T phase diagram extracted from our measurements is presented later in Fig. 4. Except for a hysteretic region which will be discussed below, we use the same symbols for the transitions extracted from up and down field sweeps, since they are essentially identical.

As the temperature is reduced, transitions 1 and 3 shift up slightly. At the lowest temperature of 1K these two transitions occur at 35.3T and 39.2T. These values are in excellent agreement with the data from [12]: 35.4T and 39.2T respectively. The middle transition shifts slightly down with temperature and then splits into two branches at about 3.7K. Both branches show hysteresis at this temperature. When the temperature reaches 1K, the lower branch of the middle transition joins with transition 1 (within the resolution of our measurements, which is $\cong$ 0.2T) and the hysteresis of this branch disappears. The second (higher) branch of the middle transition shows hysteresis in the temperature range 1-3.3K. As the temperature falls the hysteresis loop expands along the magnetic field direction, but the velocity change does not depend on the temperature. To the best of our knowledge this hysteresis has not been observed previously. The existence of hysteresis allows us to identify transition 2 as a first-order transition.

As the temperature increases above 4.5K, the position of branch 3 decreases slightly while the position of branch 2 increases. These two branches merge at 37.6T at about 7K. The position of branch 1 falls quickly and, below 25T, is similar to the behavior of the d.c. field branch observed earlier in [6-8, 11], as discussed above. At zero magnetic field, transition 1 occurs at about 17K.

Another transition is seen in the data for 7.0K and 7.8K (Fig. 2). This branch splits off of curve 1 (at about 34.5T for temperature of about 6K) and rises with temperature. At 35.8T branch 4 coincides with branch 3 for a temperature of about 8.5K. Above 10K the behavior of branch 4 (or 3) is unclear. It quickly becomes unobservable; however it is not clear whether this is associated with a small magnetoacoustic interaction at high temperatures or whether this branch moves above our upper field of 45T.

Clearly the H-T phase diagram, as measured by ultrasound in high d. c. fields, is more complicated and shows more phases than that determined with pulsed fields.

The a. c. susceptibility measurements were somewhat peripheral in this study, and hence only limited data were taken. Nevertheless, the H-T phase diagram extracted from these measurements shows a behavior similar to that of the ultrasound. At about 4K three peaks were observed in the magnetic field dependence of the a. c. susceptibility at the same fields where ultrasound velocity steps occur. While the positions of peaks 1 and 3 are nearly temperature independent in the range 1-4.5K, peak 2 shifts downward and splits into two branches when the temperature decreases below 3.9K. The lower of these two joins the line associated with transition 1, while the second displays hysteresis at 1.9K.

Above 4.5K branch 2 rises with the temperature while branch 3 falls. They coincide at 38.1T at about 7K. Branch 1 drops to zero when temperature rises to 17K. Evidence of branch 4 is also seen in the region 5.7 – 8.1K. The branches 3 and 4 coincide at a temperature of about 9.7K.

Above 9K the a. c. susceptibility signals evolve into a broad peak, the amplitude of which dramatically falls as the temperature rises. For this reason the accuracy of the branch 4 positions in this region is ≈1T and is marked by the error bars in Fig. 4. All other branches and branch 4 below 9K are measured with accuracy ≈0.2T, which is smaller than the symbol size in Fig. 4.

The phase diagram constructed above shows that the metamagnetic transition (which occurs in $URu_2Si_2$ at 35.3T in the pulsed field experiments at low temperatures) and the antiferromagnetic transition (which is observed at 17K in zero magnetic field) are two manifestations of the same transition. This result differs from the conclusion arrived at in references [6, 12], that the transitions observed in the pulsed and d. c. fields had a different origin. We suggest that this phase boundary is not seen in the pulsed field experiments due to the anomalously strong *magnetocaloric effect* (MCE), recently observed in $URu_2Si_2$ [23, 24]: a large cooling of the sample in traversing up in field through transition 1. The normal magnetocaloric effect in a paramagnet is an increase in temperature because of the magnetic field's tendency to order the spins and thereby lower the entropy. In this metal the opposite happens; the temperature drops, implying there is an increase in entropy. Figure 3 shows a trace of the magnetocaloric effect. The temperature of the sample is measured in a Si specific heat calorimeter, as the field is swept in the hybrid magnet at a rate of 2T/min. In the presence of such a temperature shift the actual temperature of the $URu_2Si_2$ sample at the transition is *lower* than the initial temperature of the helium bath (and the sample). As a result, the apparent temperature independence of the phase diagram branches measured in pulsed magnetic fields is an artifact.

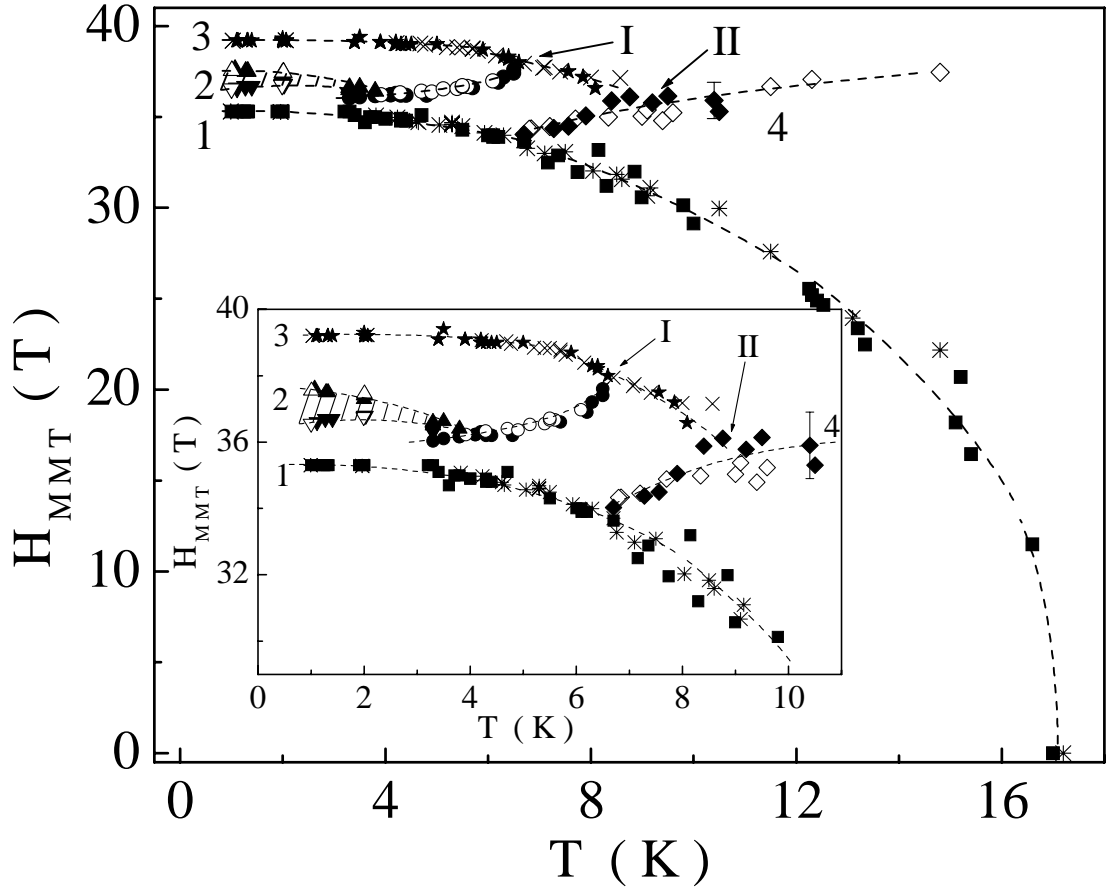

**Figure 4.** H-T phase diagram of the $URu_2Si_2$ metamagnetic transition extracted from ultrasonic (■ - branch 1; ● - branch 2, region without hysteresis; ▲ - branch 2, hysteresis region field sweeps up; ▼ - branch 2, hysteresis region field sweep down; ★ - branch 3; ◆ - branch 4) and ac susceptibility measurements (✶ - branch 1; ○ - branch 2, hysteresis region field sweep up, △ - branch 2, hysteresis region field sweep up, ▽ - branch 2, hysteresis region, field sweep down; × - branch 3; ◇ - branch 4) at continuous magnetic fields. Inset: High field/low temperature part of the same H-T diagram. The hatched area shows the hysteresis region.

It is natural to assume that point A in the pulsed field H-T diagram (see Fig. 1), where the metamagnetic transition evolves from three-steps to two-steps, and point B (Fig. 1) where the two steps merge into one, coincide to points I and II of the diagram shown in Fig. 4. In this way we can estimate the temperature change that likely occurred during the experiments in [12], for which we find 7K. We consider this value reasonable as compared to about 1K change of temperature observed during the much slower sweep rate measurements in a d. c. magnetic field (see Fig. 3). Of course, below 4.2K, where the sample is surrounded by liquid helium (or below 1.7K if the sample is placed in a He$^3$ chamber), the temperature changes in the sample will be much smaller, which is why our 1K results agree well with the results of [12] at 1.3K.

It is important to note that the phase diagram determined here is, on the whole, similar to that measured using the MCE on the *same* as well as another sample [23, 24]. The higher sensitivity of the ultrasonic technique, the smaller temperature intervals between field sweeps, and the opportunity to work below 3K, have yielded important new details of the phase diagram compared to the results obtained in [23, 24]. Nevertheless in Figure 3 of [23, 24] one can see a hint of the 4th branch in the region 6-7K.

In conclusion, the well-known transition which occurs in $URu_2Si_2$ in zero magnetic field at 17K was observed down to 1K. The field of the transition increased to 35.3T as the temperature decreases. Other phase boundaries were observed below 9K in the magnetic field region 36-40T. Hysteresis observed at low temperature for the middle transition (at about 37T) implies a first order transition. Temperature measurements were done with a Cernox resistor: magnetoresistance corrections could introduce an absolute temperature shift of up to 0.5K.

Along with the discovery of new phases in $URu_2Si_2$, two additional phenomena were uncovered in the present work. Firstly, a maximum appears at 29T in the magnetic field dependence of the ultrasonic velocity for which there is no explanation at present. We assume that this maximum and the maxima observed in magnetoresistance [15] and Hall coefficient [16] of $URu_2Si_2$ at 29T have the same yet-unknown origin. Secondly, the behavior of velocity at the 39T transition is very unusual: at temperatures above about 4K the velocity decreases discontinuously as the magnetic field rises; at 4K there is no evidence of a transition; and below 4K the velocity increases discontinuously. We assume that this transition is not associated with lattice properties of $URu_2Si_2$ but more likely relates to ordering of electronic system or to spin density waves, for example.

We are grateful to D. Hall for help with the a. c. susceptibility measurements and to D. Agterberg for useful discussions. We would also like to thank M. Jaime for introducing us to the magnetocaloric effect measurements. This research was supported by the NSF grant DMR-9971123, the measurements were performed at the Tallahassee site of the National High Magnetic Field Laboratory, which is supported by the State of Florida and the U.S. Department of Energy.